# Persistent anisotropy of the spin cycloid in BiFeO$_3$ through ferroelectric switching

## Authors

Peter Meisenheimer*,[1], Guy Moore[1,2], Shiyu Zhou[3], Hongrui Zhang[1], Xiaoxi Huang[1], Sajid Husain[1,2], Xianzhe Chen[1,2], Lane W. Martin[1,2,4], Kristin A. Persson[1,5], Sinéad Griffin[2,5], Lucas Caretta[3,6], Paul Stevenson*,[7], Ramamoorthy Ramesh[4,8,9]

## Affiliations

[1] Department of Materials Science and Engineering, University of California, Berkeley, CA, USA

[2] Materials Sciences Division, Lawrence Berkeley National Laboratory, Berkeley, CA, USA

[3] Department of Physics, Brown University, Providence, RI, USA

[4] Department of Materials Science and NanoEngineering, Rice University, Houston, TX, USA

[5] Molecular Foundry, Lawrence Berkeley National Laboratory, Berkeley, CA, USA

[6] School of Engineering, Brown University, Providence, RI, USA

[7] Department of Physics, Northeastern University, Boston, MA, USA

[8] Department of Physics, University of California, Berkeley, CA, USA

[9] Department of Physics and Astronomy, Rice University, Houston, TX, USA

* corresponding authors; meisep@berkeley.edu, p.stevenson@northeastern.edu

## Abstract

A key challenge in antiferromagnetic spintronics is the control of spin configuration on nanometer scales applicable to solid-state technologies. Bismuth ferrite (BiFeO$_3$) is a multiferroic material that exhibits both ferroelectricity and canted antiferromagnetism at room temperature, making it a unique candidate in the development of electric-field controllable magnetic devices. The magnetic moments in BiFeO$_3$ are arranged into a spin cycloid, resulting in unique magnetic properties which are tied to the ferroelectric order. Previous understanding of this coupling has relied on average, mesoscale measurements to infer behavior. Using nitrogen vacancy-based diamond magnetometry, we show that the spin cycloid can be deterministically controlled with an electric field. The energy landscape of the cycloid is shaped by both the ferroelectric degree of freedom and strain-induced anisotropy, restricting the magnetization changes to specific ferroelectric switching events. This study provides understanding of the antiferromagnetic texture in BiFeO$_3$ and paves new avenues for designing magnetic textures and spintronic devices.

**Introduction**

In the push to realize spintronic devices for computation, controlling magnetism using electric fields at the device scale remains an elusive challenge[1,2]. The multiferroic BiFeO$_3$ (BFO) exhibits both room-temperature large ferroelectric polarization and canted antiferromagnetism with deterministic coupling between these order parameters through the crystal symmetry[1–6]. Thus, the magnetic nature of BFO can be controlled with electric fields, potentially paving the way for ultra-energy-efficient magnetic and spintronic devices with more favorable scaling[6–10]. Indeed, electric-field control of magnetism in coupled ferromagnetic layers using BFO been the subject of considerable attention[4,11–13], and while this functionality has long excited researchers, understanding of the process has generally relied on mesoscale imaging and transport measurements to infer the structure and behavior of the BFO. This has left gaps in our understanding of the microscopic mechanism and details of this process.

Only recently has the opportunity to directly image and understand complex magnetic structures at their native scales become possible using techniques such as nitrogen-vacancy (NV) diamond-based scanning probe magnetometry (henceforth NV microscopy)[14,15]. While recent works have shed new light onto the nanoscopic magnetic structure of BFO[16,17], a significant fundamental question remains as to how the complex magnetism interacts with the ferroelectric order and electric fields. In BFO, ferroelectric switching pathways are complex[11,18,19], proceeding through coupled, multi-step rotations of the polarization. The corresponding magnetic-switching phenomena can be tied to the ferroelectric polarization evolution[4,12,20], but observation of these coupled phenomena in real space on the nanometer scale has not been possible. Here, using NV microscopy, we directly image the stray magnetic field at the surface resulting from the spin cycloid as it couples to ferroelectric domains and complex (71°, 109°, and 180°) ferroelastic and ferroelectric switching events.

In the bulk, BFO exhibits an antiferromagnetic (AFM) spin cycloid that exists in the plane defined by $\boldsymbol{P}$ and the propagation direction, $\boldsymbol{k}$, which points along a ⟨110⟩ that is orthogonal to $\boldsymbol{P}$ [21–25]. This vector connects second-nearest neighbor iron sites which, in an unperturbed G-type AFM, would be ferromagnetically coupled within a {111} [25–27]. The spin cycloid itself has been modeled as a Néel-type, rotating uncompensated magnetization, $M(\boldsymbol{r})$, that exists in the plane defined by $\boldsymbol{k}$ and $\boldsymbol{P}$ (i.e., the (11$\bar{2}$), where $\boldsymbol{P}$ is along the [111] and $\boldsymbol{k}$ along the [$\bar{1}$10], unless otherwise noted) with a period of ~65 nm (**Figure 1a**). This has been described as

$$M(r) = m[\cos(k \cdot r) \, e_k + \sin(k \cdot r) \, e_p], \quad (1)$$

where $m$ is the volume-averaged magnetization, $|k| = 2\pi/\lambda$, $r$ is a coordinate in real space, and $e_k$ and $e_P$ are the unit vectors in the directions of $k$ and $P$, respectively[14]. A second-order canting also exists due to the Dzyaloshinskii-Moriya interaction (DMI) arising from the antiferrodistortive octahedral rotations[26,28–30], causing the magnetization to buckle slightly out of the $k$-$P$ plane (**Figure 1b**). This second-order spin-density wave, noted here as $M_{SDW}$, can be described[16] by

$$M_{SDW}(r) = m_{DM} \cos(k \cdot r) \, (e_k \times e_p). \quad (2)$$

Example solutions to eqs. 1 and 2 are provided in **Supp. Figure 1**. The direction of $k$, and the magnetization are thus intimately tied to the ferroelectric polarization of BFO, consistent with the fact that the spontaneous polarization is the primary order parameter of multiferroicity in this system.

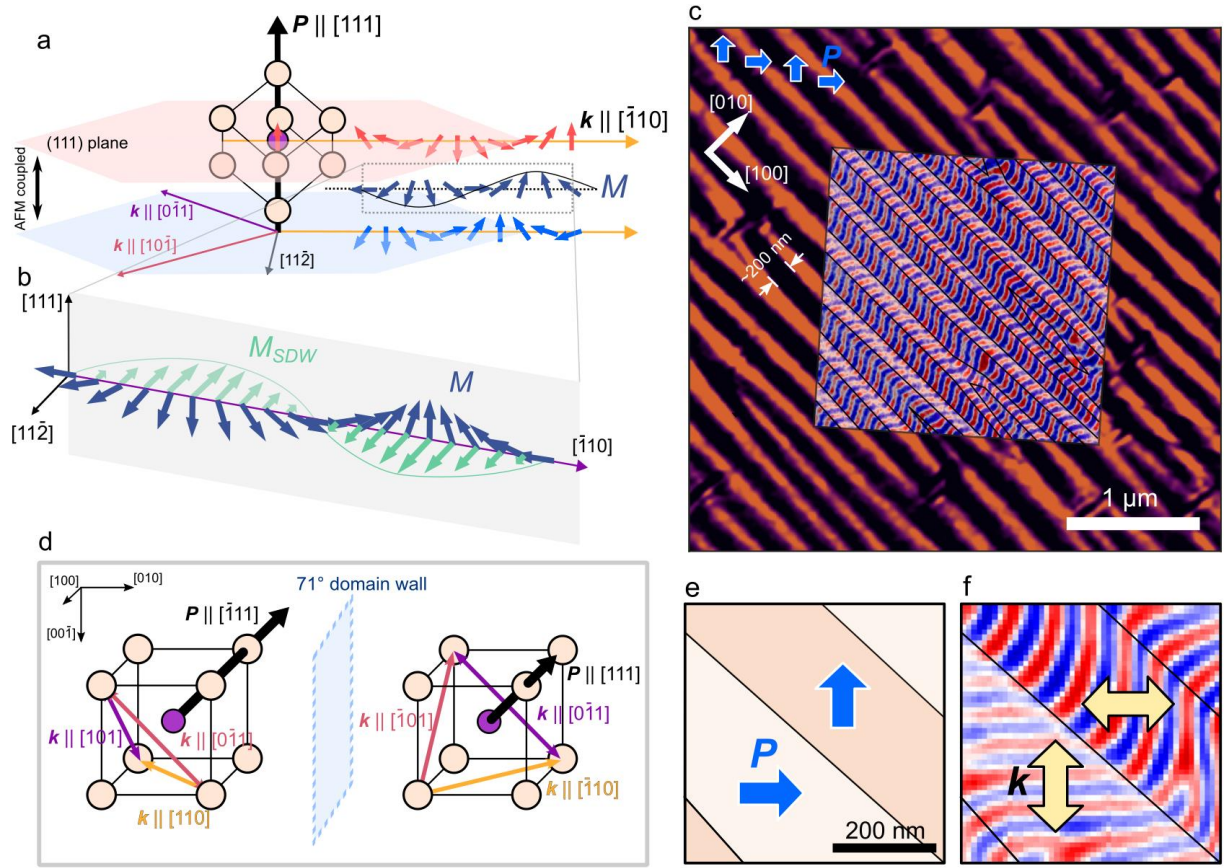

**Figure 1 | Spin cycloid in BiFeO$_3$. a** Iron moments in BFO are antiferromagnetically aligned along the [111], modulated by the cycloid propagation along $k$, [$\bar{1}$10]. Other allowed directions of $k$ also lie within this (111). The canting of the AFM alignment gives rise to an uncompensated magnetization, $M(r)$, which rotates primarily in the $k$-$P$ plane with the same period as the antiferromagnetic moments, ~65 nm. **b** $M$ is

further frustrated by DMI associated with the octahedral rotations, giving rise to a modulation $M_{SDW}(r)$ out of the $k$-$P$, $(11\bar{2})$ plane. The $(11\bar{2})$ plane is shown by the shaded plane and $M_{SDW}$ points along the $[11\bar{2}]$ direction. **c** From PFM, BFO samples show stripe 71° ferroelectric domains along [100] that are characteristic of the material. NV images taken at the same location show that $B(r)$ and $P(r)$ can be exactly mapped where $k \perp P$. **d** Illustration showing the allowed directions of $k$ and the change across the 71° ferroelectric domain wall. Because the in-plane variant of $k$ (gold) is favored and this is tied to the direction of $P$, these domain walls in BFO give rise to a 90° rotation in $k$. This is illustrated in **e,f**, where the local $P$ can be mapped to $B(r)$ with continuous reorientations occurring every ~200 nm at domain walls.

In NV microscopy, the stray magnetic fields ($B(r)$) from the sample surface perturb the energy states of a single NV center implanted into a diamond scanning probe tip. By measuring the optically detected magnetic resonance spectra of the NV center, one can detect small magnetic fields (<2 $\mu T \sqrt{Hz}^{-1}$) with spatial resolution down to at least 10 nm. Combined with piezoresponse force microscopy (PFM), it is possible to locally map the vectorial polarization distribution, $P(r)$, in the ferroelectric domains and thereby correlate the relationship between the magnetic and ferroelectric structure in BFO. While the interaction of the cycloid with ferroelectric domains has been suggested in previous work, where poling areas of the film results in a locally uniform $k$ (the cycloid propagation direction), the interaction between ferroelectric domain walls and the cycloid has not been directly demonstrated[14]. More importantly, it is not well understood if and how the cycloid propagation direction changes during ferroelectric switching, a question especially relevant to electric-field manipulation of magnon transport[20,31] and exchange coupling across heterointerfaces[4,11]. Here, it is shown that direct mapping of the canted antiferromagnetic texture to the ferroelectric domains can be achieved. Of greater importance[20,31], it is shown that upon applying an electric field to switch the ferroelectric polarization, the relationship between $k$ and $P$ is conserved, with $k$ showing a strong anisotropy along the in-plane [110] and $[\bar{1}10]$ which are both perpendicular to $P$. It is also observed that this anisotropy persists agnostic to the direction of electric field, through both in-plane and out-of-plane ferroelectric switching events.

**Anisotropy of the cycloid propagation**

The model heterostructures studied here are ~100 nm thick, (001)-oriented BFO thin films deposited on $DyScO_3$ (DSO) (110) substrates using pulsed-laser deposition, both with and without metallic $SrRuO_3$ bottom electrodes (Methods). In both configurations (with and without bottom electrodes), X-ray diffraction confirms that heterostructures are constrained by the substrate (**Supp. Figure 2**). Subsequent PFM imaging reveals 71° stripe-like ferroelectric domains (**Figure**

**1c**); a commonly observed characteristic[32] of BFO. In turn, NV microscopy of the same shows the ~65 nm period sinusoidal modulation of magnetization due to the spin cycloid[14,17]. By carrying out PFM and NV microscopy at the same location, we observe that the 71° ferroelectric domain walls match exactly with changes in the cycloid propagation, switching from [110] to [$\bar{1}$10] (horizontal and vertical in **Figure 1c-f**) with changes in $P$. This behavior is due to the magnetoelectric coupling between $k$ and $P$ and demonstrates a nondegeneracy in the cycloid landscape. Generally, because $k$ must remain orthogonal to $P$, when $P$ is along [$\bar{1}$11], $k$ can propagate along [110], but when $P$ changes across a domain wall to [111], $k$ must reorient to an allowed direction orthogonal to the new $P$, either [$\bar{1}$10], [$\bar{1}$01], or [0$\bar{1}$1] (**Figure 1d**). In bulk BFO, all three of these variants are allowed[16]. In the thin film heterostructures studied here, however, $k$ appears to select only the purely in-plane directions, [110] or [$\bar{1}$10], switching every ~200 nm commensurate with the ferroelectric domain walls (**Figure 1e,f**).

In the as-grown configuration, $k$ is observed to be along both the [110] and [$\bar{1}$10], in each case selecting the allowed direction without an out-of-plane component dependent on the ferroelectric polarization. We hypothesize that the preference for the in-plane $k$ vectors arises from epitaxial constraint imposed on the BFO film from the DSO substrates (e.g. differences in symmetry, lattice constant, etc.). It has been previously observed that, in thin-film BFO, the relationship between $k$ and $P$ can be controlled through the choice of substrate and, at small compressive strains, the cycloid follows similar behavior to bulk[14,16,24], where $k$ is bound to a ⟨110⟩ that is orthogonal to $P$ (so-called type-I). At higher epitaxial strains, however, a type-II cycloid is observed, which propagates along a ⟨11$\bar{2}$⟩ perpendicular to $P$[16,17,33]. It appears, then, that while constraints imposed by the DSO substrate allow for the bulk-like, type-I cycloid, it creates a strongly anisotropic landscape which inhibits out-of-plane projections of $k$. This effect on the anisotropy of $k$, however, has not been determined. To better understand the impact of this anisotropy on the spin texture, density functional theory (DFT) calculations have been performed to explore the canted magnetic moment in BFO.

While DFT simulations can provide powerful insight into the local atomic structure, predicting the ground-state magnetic order of BFO is computationally intractable using plane-wave DFT due to the long period of the cycloid (~65 nm) and the corresponding large system size required to fully simulate it[34]. A practical alternative is to use the generalized Bloch condition for $q$-spirals which can account for rotations in the cycloidal plane, but this cannot predict moments that are canted out of this plane due to boundary conditions. To then help understand the ground-state magnetization of the system, here we discretize the magnetic structure into subsections manageable by first principles. Starting from a 2x2x2 unit cell with G-type antiferromagnetism, where anti-aligned iron moments point along the vector $\pm L$, we systematically rotate the initial $L$ within a fixed plane (given by the angle $\phi$, **Figure 2a**) to simulate the cycloid and resolve the canted magnetism. In this case, $\phi$ is in the $(11\bar{2})$, defined by $P$ and $k$, which hosts the cycloid. Additionally, calculations are performed applying on-site Hubbard $U$ and Hund $J$ corrections to the O-$p$ manifold, in addition to the Fe-$d$ and Bi-$p$, with further detail in **Methods**. We first demonstrate the validity of our approximations and methodology by reproducing the established spin texture, $L$, as reported in the literature. Specifically, we rotate $L$ within an orthogonal plane, (111) which leads to an approximately two-times greater energy cost than $L$ in $(11\bar{2})$, relative to the reference minimum energy spin quantization axis (**Figure 2b).**

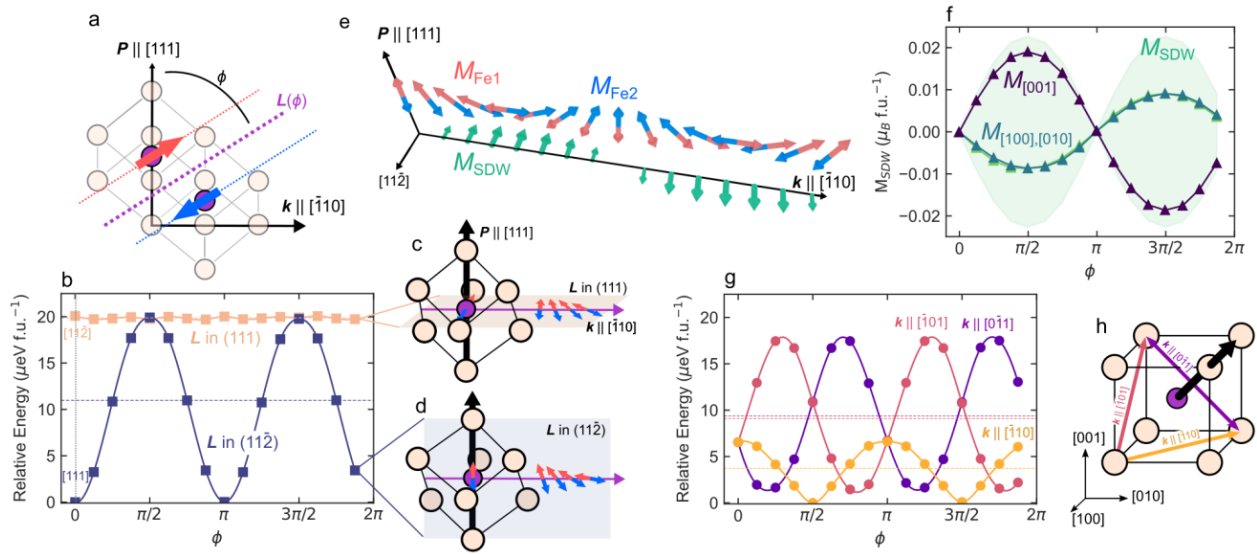

**Figure 2 | Resolution of M$_{SDW}$ from first principles. a** Schematic showing the initialization angle $\theta$ within the cycloid plane for the antiferromagnetically aligned Fe spins. **b** Comparison of the magnetocrystalline anisotropy when the Fe spins are rotated in the (111) and $(11\bar{2})$. This agrees with the expectation that the cycloid rotates within $(11\bar{2})$, as the mean value of the energy is 2x higher when moving the rotation to the (111) plane. **c,d** Illustrations of the rotation planes in b. **e** Graphical representation of the magnetic cycloid directly from DFT, showing the relaxed Fe spins (red, blue), as well as the resulting canted moment (green), $M_{SDW}$, that spontaneously arises. **f** Components of the relaxed canted moment, showing the periodicity of M$_{SDW}$ which follows the cycloid period. The components sum to a vector of varying magnitude along $[11\bar{2}]$ and $M_{SDW}$ is largest when the Fe spins are pointed along $[\bar{1}10]$. **g** Relative energy along the three possible $k$ directions when the unit cell is epitaxially strained to DSO. The mean energy is 2x lower when the cycloid

propagates along the in-plane $[\bar{1}10]$ direction, agreeing with our experimental observation. The dotted lines show the mean energy values. **h** Schematic of the three $k$ directions in g.

Initializing the iron moments along the rotation angle $\phi$ in the $(11\bar{2})$, the canted $M_{SDW}$ component of magnetization along the $[11\bar{2}]$ spontaneously arises when the structure is relaxed, shown graphically in **Figure 2e** and quantitatively in **Figure 2f**. In these simulations, $M_{SDW}$ is the net moment that comes from the canting of the atomic moments away from the initialization direction $L$ and is reported as the vector sum of the iron spins. From our simulations, $M_{SDW}$ follows the same period as the cycloid and reaches a maximum value when $L$ is parallel to $[\bar{1}10]$. This is consistent with the expectation from symmetry that $M_{SDW}$ emerges due to the DMI from octahedral rotations with their axis along the polarization direction[26,29,30], where $\boldsymbol{D}_{ij} \cdot (\boldsymbol{S}_i \times \boldsymbol{S}_j)$ is maximized when $\boldsymbol{D}_{ij}$ and $\boldsymbol{S}_{i,j}$ are orthogonal, in this case $\boldsymbol{S}_{i,j} \parallel \boldsymbol{k}$. The value of $M_{SDW}$ reaches a maximum of 0.02 $\mu_B$, which is consistent with previous predictions[34,35] and is of the order of previous experimental results[14,16,22,29]. Discretizing the magnetic cycloid in this way then produces results that agree exactly with previous experimental and theoretical interpretations of the canted $M_{SDW}$ component of the cycloid[28–30]. Additionally, reducing the ferroelectric polarization or rotating the cycloid in another plane removes this phenomenon (**Supp. Figure 3**), showcasing its physical origin.

With the magnetic structure reproduced using DFT, we next consider how the anisotropy introduced into the system through epitaxial constraints. Fixing the unit cell to the in-plane lattice constants of DSO, a similar calculation is performed to that above where the iron moments are rotated in the planes defined by $P$ and the three possible $k$ directions (i.e., $[\bar{1}10]$, $[\bar{1}01]$, and $[0\bar{1}1]$). From these data, there is a clear anisotropy favoring the $k \parallel [\bar{1}10]$ due to epitaxy, with a mean energy of approximately two-times lower than the other symmetry allowed $k$, in line with the experimental results. This is in contrast to the bulk, zero-strain state, where all possible directions of $k$ are symmetry and energetically equivalent (**Supp. Figure S5**). Though, as mentioned above, we cannot accurately simulate the interaction between neighboring iron moments along $k$ that give an additional potential energy gain due to the cycloid[26], it is noteworthy that in this colinear limit of the antiferromagnetic structure, we would predict a ground state with $L$ pointing along $[\bar{1}10]$ with $P$ along $[111]$; the same as suggested previously[11]. Understanding, then, that substrate constraints intrinsically break the degeneracy of the allowed $k$ directions, one can then ask whether the state of the cycloid can be deterministically changed with an electric field and whether this anisotropy is strong enough to persist in the switched state.

**In-plane electric field switching of the cycloid**

Through application of an in-plane electric field, perpendicular to the striped 71° domain walls, we can reorient the in-plane component of polarization in BFO to produce a net polarization ($P_{net}$) along the $[100]$ or $[\bar{1}00]$, composed of ($[11\bar{1}]$ and $[1\bar{1}\bar{1}]$) or ($[\bar{1}1\bar{1}]$ and $[\bar{1}\bar{1}\bar{1}]$) polarized domains, respectively. Test structures (**Methods**) used to apply this in-plane field and the corresponding ferroelectric switching behavior are shown (**Figure 3a,b**). The ferroelectric domains measured by PFM are shown for devices, respectively poled into the $P_{net} \parallel [010]$ (**Figure 3c**) and $P_{net} \parallel [0\bar{1}0]$ (**Figure 3d**) configurations. Mapping the spin cycloid, measured through NV microscopy, to these domain images, the changes in $k$ map directly to the ferroelectric domain walls and the sense of the cycloid such that the relationship $k$ perpendicular to $P$ is preserved. Here, for example, $[11\bar{1}]$ and $[\bar{1}\bar{1}\bar{1}]$ polarized domains show equivalent $k$ axes along $[\bar{1}10]$.

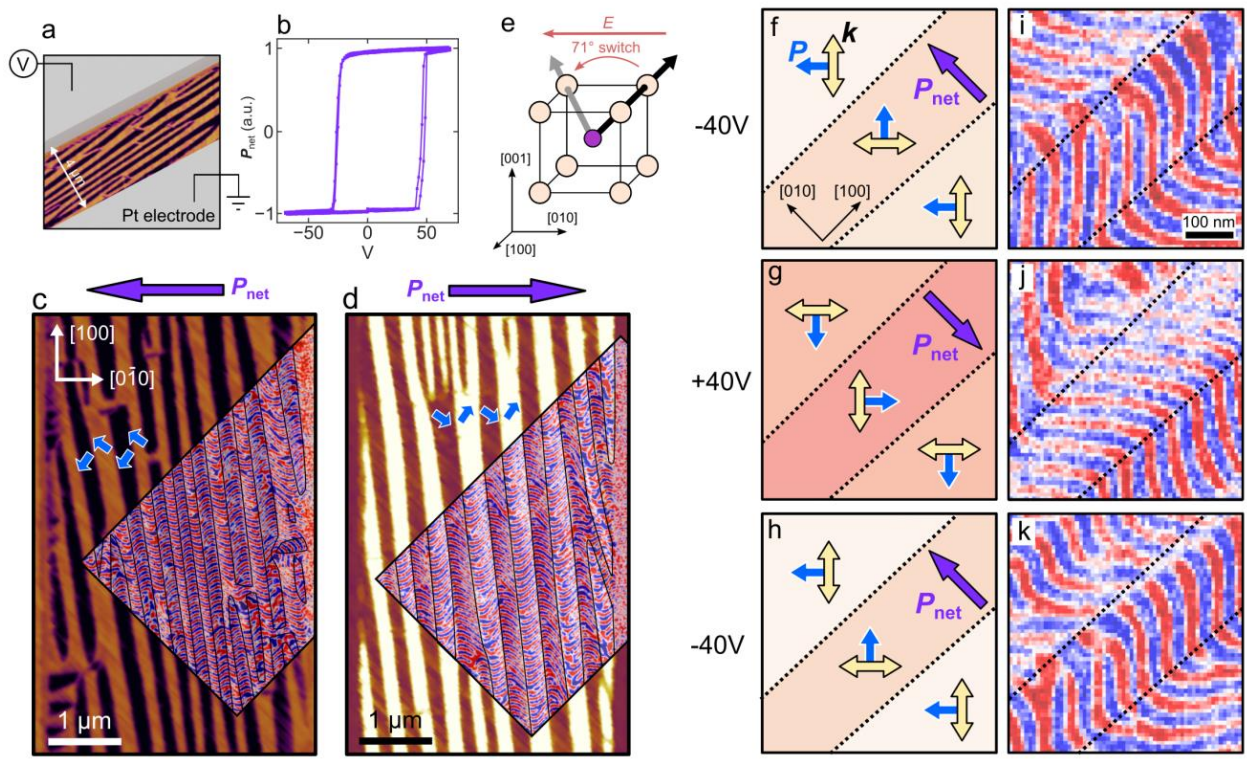

**Figure 3 | In-plane switching of ferroelectric domains. a** Schematic of the test structure used to apply an in-plane electric field along the [010] direction, showing the orientation of the ferroelectric domain walls perpendicular to the applied field. **b** Ferroelectric hysteresis loop of the device in a, where ferroelectric switching happens at -30/+40 V. **c,d** Ferroelectric domains overlayed with NV magnetometry data for devices poled both with net polarization left and right. Ferroelectric domain walls are shown as black lines

on the NV data. **e** illustration showing the 71° in-plane ferroelectric switch when the electric field is applied along [0$\bar{1}$0]. **f-h** Illustrations showing the 71° ferroelectric domain variants when poled along [010] (purple arrow). The in-plane projections of ***P*** and ***k*** are shown in each domain with blue and yellow arrows. **i-k** In situ NV images which map to the ferroelectric domains in b,c,d. Here the 90° change in the directionality of ***k***, from [110] to [$\bar{1}$10] (and vice versa) can be seen clearly between switching events.

As in the case of the as-deposited films, ***k*** remains parallel to the surface of the film. If the polarization of a domain reorients from [111] to [$\bar{1}$11], ***k*** selects the [110] out of the three symmetry allowed axes. We believe that this may be due to the biaxial strain state imposed by the DSO substrate, as this strong in-plane anisotropy is not observed in $BiFeO_3$ deposited on other $REScO_3$ substrates[16,17], such as $GdScO_3$, where the ***k*** || [011] propagation appears to be favored. With this assumption, that ***k*** is confined to the in-plane directions and we will observe a 90° change in ***k*** under 71° ferroelectric switching, we measure the change in ***k*** *in situ* at a single location under electric field.

With an in-plane electric field, individual ferroelastic domain walls will tend to remain stationary and the polarization of individual domains reorients by 71° (**Figure 3e**)[4,11,20]. The stray magnetic field measured at a single location *in situ* and the corresponding ferroelectric domains are shown (**Figure 3f-k**) where, e.g., in the center domain, the directionality of ***k*** changes from [110] to [$\bar{1}$10] and back to [110] under successive switching events. Again, after ferroelectric switching, the relationship between ***k*** and ***P*** is conserved, with both rotating 90° about the [001]-axis. In the case of BFO, this observation is not necessarily surprising. In previous works, electrical switching of the polarization in BFO has been proposed to follow successive rotations of ***P*** (i.e., ferroelastic switching) instead of through an intermediate state which is a nonpolar configuration (i.e., ferroelectric switching)[11,18]. From this framework of the deterministic rotation of ***P*** and the $FeO_6$ octahedra, the case of an in-plane field becomes easy to understand: if ***P*** rotates 90° about the [001] (as in a 71° switch), and ***P*** ⊥ ***k***, we would expect that ***k*** will also rotate correspondingly by 90° about the [001].

**Out-of-plane electric field switching of the cycloid**

In the case of an electric field applied along the [001], however, this switching pathway becomes more complex. Using a BFO heterostructure deposited with a $SrRuO_3$ back electrode, we can explore how electric fields in the out-of-plane direction, and more complex ferroelectric switching events, influence the reorientation of the spin cycloid. An out-of-plane ([001]-oriented) electric field was applied locally to the sample using the PFM tip as the top electrode while grounding the bottom SrRuO$_3$ layer (Methods). A voltage of +8 V was applied to the heterostructure within a ~5 μm area, and -8 V was applied to a ~2.5 μm area within the previously switched region to return it to the same $P_{net}$ as the as-grown state (a so-called box-in-a-box structure). PFM is then used to measure the switched area at multiple scan angles to vectorize the polarization (Methods and **Figure 4a,b**). Using the polarization vectors and calculating the angle between them before and after switching allows us to construct a map of ferroelectric switching events (**Figure 4c**). This data can then be compared to NV magnetometry measurements to study the dependence of the cycloid propagation direction on local switching events.

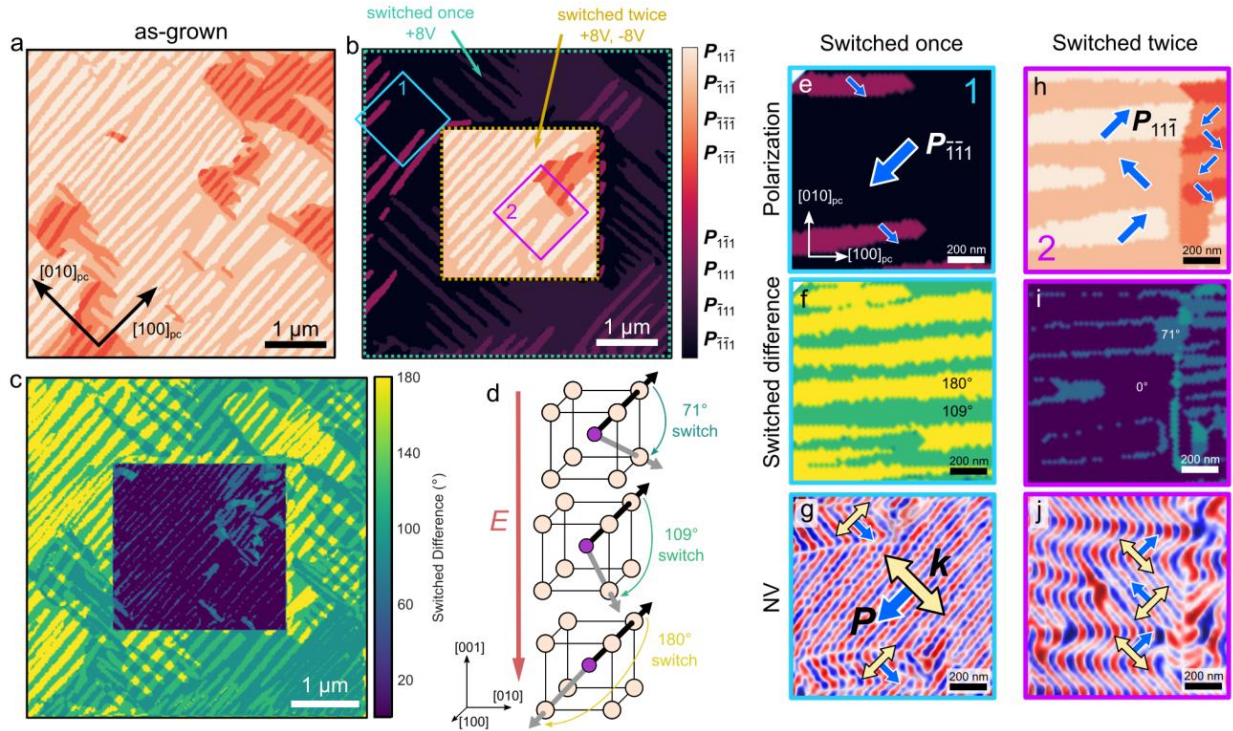

**Figure 4 | Out-of-plane switching of the ferroelectric domains. a,b** Vector map constructed from the PFM of BiFeO$_3$ samples in the (a) as-deposited and (b) out-of-plane poled configurations. The sample is poled using the PFM tip as the top electrode and +/-8V, corresponding to an electric field of ~2 MV cm$^{-1}$. **c** Map of the reorientation of ferroelectric domains, calculated from the difference of the polarization vectors in a and b. **d** Illustration showing the different ferroelectric switching events under the application of an out-of-plane electric field. **e,h** zoomed-in areas of the ferroelectric polarization, shown by the blue and purple boxes labeled 1 and 2 in b. The in-plane projection of the ferroelectric polarization is shown by the blue arrows. **f,i** Calculated switching events from c at the same locations as e,h. **g,j** NV microscopy images showing the spin cycloid at the same locations in e,h. $P$ and $k$ are shown by blue and yellow arrows.

The map of the ferroelectric polarization direction (**Figure 4a,b**), the difference between the as-grown and switched states (**Figure 4c**), and the local magnetization in both the singly (**Figure 4e-g**) and doubly poled (**Figure 4h-j**) regions are shown. It can be seen (**Figures 4f,i**) that there are 71°, 109°, and 180° ferroelectric switching events during the poling process. As in the as-grown material, $k$ is confined to the [110] and [$\bar{1}$10] within the (001) and is oriented perpendicular to the final state of $P$, regardless of the local switching event being a 71°, 109°, or 180° rotation of the polarization. This is true for both the singly and doubly switched areas. The interpretation is then that $k$ is dominated by the magnetoelastic energy from the substrate constraints and the final polarization vector direction.

Previous reports on BFO have demonstrated that, when subjected to an out-of-plane oriented electric field, $P$ will not rotate continuously through 180° or through a nonpolar intermediate state, but switching will be accomplished by successive 71° and 109° switching events[11,18]. The data here mirror these observations, but the anisotropy of $k$ is showcased by the fact that the highest symmetry operations to rotate $P$ do not preserve the experimentally observed directions of $k$. These potential pathways and the operations to preserve the directionality of $k$ are shown (**Supp. Figure S8**). In other words, if $k$ were not anisotropic in the (HK0), it would be expected to change with the highest symmetry operation of $P$. This observation mirrors previous results on the deterministic nature of the switching in BFO[11], where switching only proceeds along experimentally observed pathways if the antiferromagnetic order parameter, $L$, is fixed to the ⟨110⟩ in the (001), the same as $k$ observed here.

Here we show that the antiferromagnetic spin cycloid in BFO is intimately tied to the direction of polarization, even after ferroelectric switching events. In BFO thin films deposited on DSO, the propagation vector of the cycloid, $k$, is confined to directions orthogonal to $P$ with no out-of-plane component, [110] and [$\bar{1}$10]. Discontinuities in the cycloid propagation map directly to 71° ferroelectric domain walls, where $k$ must reorient to maintain this relationship. As the magnetic structure can be not only tuned in the ground state, but the functionality can be constrained in-plane using the substrate (e.g., symmetry, lattice mismatch), this may provide a method for designing particular material architectures to optimize spin and magnon transport[20,31]. The fixed in-plane anisotropy of the antiferromagnetic cycloid is especially important for the implementation of vertically oriented spintronic devices[9,10]. Additionally, the ability to potentially engineer local frustration in the magnetic structure may open the door for electrical creation of magnetic topological defects at particular domain wall interactions[36–39], as confinement-induced frustration is demonstrated to be a powerful method for engineering complex textures in ferroics[40–42].

## Methods

### Sample Fabrication

BFO thin films were deposited on DSO substrates at 700 °C using pulsed laser deposition with a laser energy density of ~1.5 J cm$^{-2}$ in a background oxygen pressure of 90 mTorr. X-ray diffraction was measured on a Panalytical diffractometer with CuK$\alpha$ source. Test structures were patterned using standard lithography techniques.

Ferroelectric domains were mapped using piezoresponse force microscopy in an asylum MFP-3D, both in-plane and out-of-plane, at two orthogonal sample rotations to vectorize the polarization.

### NV microscopy

BFO samples were measured at room temperature using a commercial scanning NV microscope (Qnami ProteusQ) which combines a confocal optical microscope with a tuning-fork based atomic force microscope. Diamond tips with a parabolic taper containing single NV centers were used to increase photon collection efficiency (Quantilever MX+). The orientation of the NV center in the lab frame was determined using the method outlined in [43] and a Ta(2nm)/MgO/CoFeB(0.9nm)/Ta(5nm) sample with perpendicular magnetic anisotropy.

The magnetic field is quantitatively determined by measuring the optically detected magnetic resonance spectrum of the NV center at each point in space, yielding the projection of the magnetic field along the NV center axis. Large area images were collected in the qualitative "dual iso-B" mode, where the response of the NV center to two different microwave frequencies is used to track the magnetic field, as detailed in Ref [44]. This dramatically reduces data acquisition time with respect to collecting the full optically detected magnetic resonance spectrum.

### Electronic Structure Calculation Details

DFT calculations were carried out in VASP. Hubbard U and Hund J corrections for the valence states of all species were calculated using collinear DFT using a linear response (LR) workflow developed in the atomate code framework[45]. Linear response analysis was performed without the inclusion of SOC to reduce computational cost, and because SOC has been found in other systems to have a relatively small effect on the on-site corrections from LR[45]. These values were calculated to be *U*, *J* = 5.2, 0.4 eV for Fe-d, *U*, *J* = 0.8, 0.8 eV for Bi-p, and *U*, *J* = 9.7, 1.9 eV for O-p. Because MAE is on the order of µeV, geometry relaxation, spin-orbit coupling (SOC), and on-site Hubbard *U* and Hund *J* corrections are all accounted for in a holistic manner to address their interdependence. In all calculations, Hubbard and Hund corrections are applied to all outer shell manifolds (Fe-*d*, Bi-*p*, and O-*p*). These structures were calculated for the structure endpoint reported in Ref. [18]. Using these calculated on-site Hubbard corrections, full geometry relaxation of the structure was performed with SOC included until self-consistency was reached for an electronic energy tolerance of $10^{-6}$ eV. All calculations were performed for the 2x2x2 supercell (eight times the formula unit) to accommodate the G-type antiferromagnetic structure. These computational subtleties are addressed in greater depth in the **Supp. Figure S4 and Note S1**.

**Acknowledgements**

This work was primarily supported by the U.S. Department of Energy, Office of Science, Office of Basic Energy Sciences, Materials Sciences and Engineering Division under Contract No. DE-AC02-05-CH11231 (Codesign of Ultra-Low-Voltage Beyond CMOS Microelectronics (MicroelecLBLRamesh)) for the development of materials for low-power microelectronics. P.M., L.W.M., and R.R. additionally acknowledge funding from the Army Research Office under the ETHOS MURI via cooperative agreement W911NF-21-2-0162. S.Z. and L.C. acknowledge funding from Brown School of Engineering and Office of the Provost. P.S. acknowledges support from the Massachusetts Technology Collaborative, Award number #22032. G.M. acknowledges support from the Department of Energy Computational Science Graduate Fellowship (DOE CSGF) under grant DE-SC0020347. Computations in this paper were performed using resources of the National Energy Research Scientific Computing Center (NERSC), a U.S. Department of Energy Office of Science User Facility operated under contract no. DE-AC02-05CH11231. The work performed at the Molecular Foundry was supported by the Office of Science, Office of Basic Energy Sciences, of the U.S. Department of Energy under the same contract under Contract No. DEAC02-05CH11231.

**Author Contributions**

P.M., P.S., L.C., and R.R. designed experiments. H.Z., and X.H. synthesized and patterned samples. P.M., P.S., and S.Z. performed NV microscopy. P.M. performed PFM measurements. G.N. and S.G. performed DFT calculations. P.M., P.S., L.C., and R.R. analyzed results and wrote the paper. All authors have contributed to manuscript revisions.

**Competing interests**

The Authors declare that they have no competing interests

**Data and materials availability**

The PFM and NV data generated in this study have been deposited to Zenodo under DOI 10.5281/zenodo.8310434. Other data used in these experiments are available from the authors upon reasonable request.

**Extended Data**

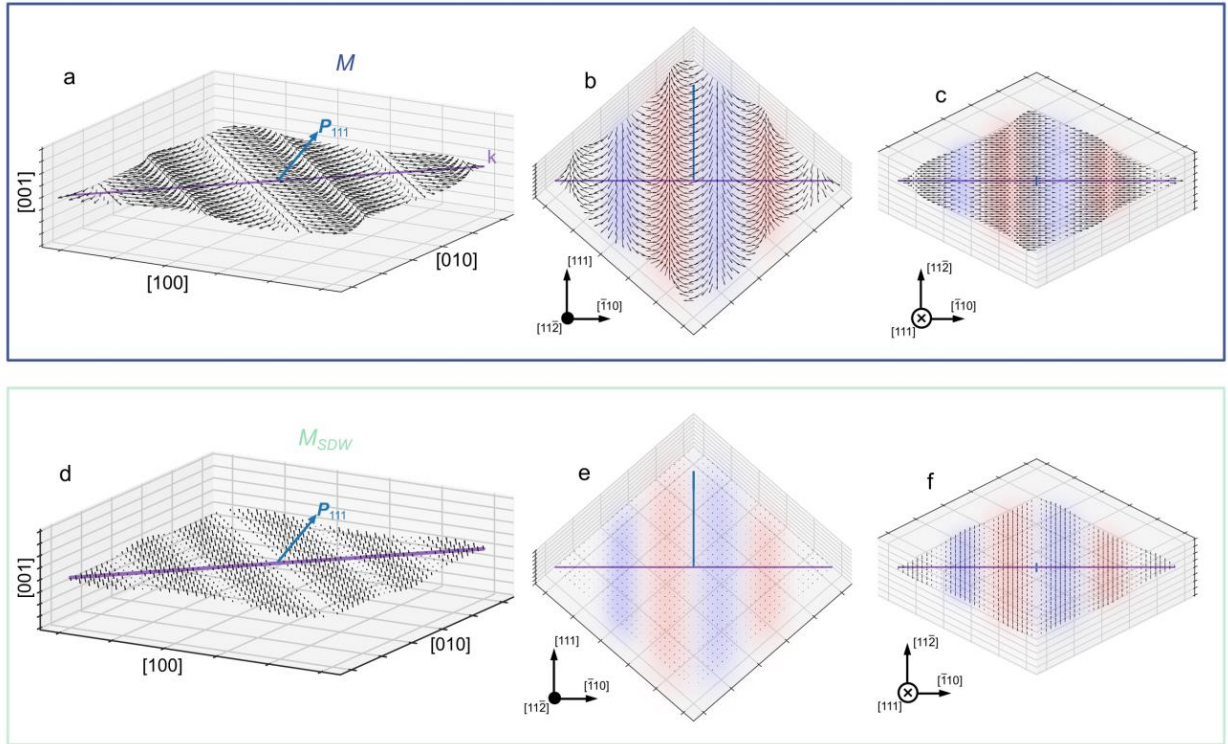

**Supp. Figure S1 | 3D representation of the spin cycloid.** Vector plots of solutions to **Eqs. 1** and **2**, dictating empirical descriptions of $M$ (**a-c**) and $M_{SDW}$ (**d-f**). Views along $[11\bar{2}]$ and $[111]$ ($\perp$ and $\parallel$ to $P$) are shown, illustrating that $M$ rotates in the $(11\bar{2})$, defined by $P$ and $k$, and $M_{SDW}$ exists in the $(111)$ along the $[11\bar{2}]$. The red and blue shading map to the images measured via NV magnetometry. Previous works[14,16] propose that the signal observed in NV magnetometry is due primarily or completely to the component of $M_{SDW}$, shown here in the $[11\bar{2}]$ direction (**f**). As described in the text, $M_{SDW}$ is largest when the Fe spins point along $[\bar{1}10]$; as $M$ is always perpendicular to the axis of the Fe spins, $M_{SDW}$ is largest when $M$ points along $[111]$, shown in c and f.

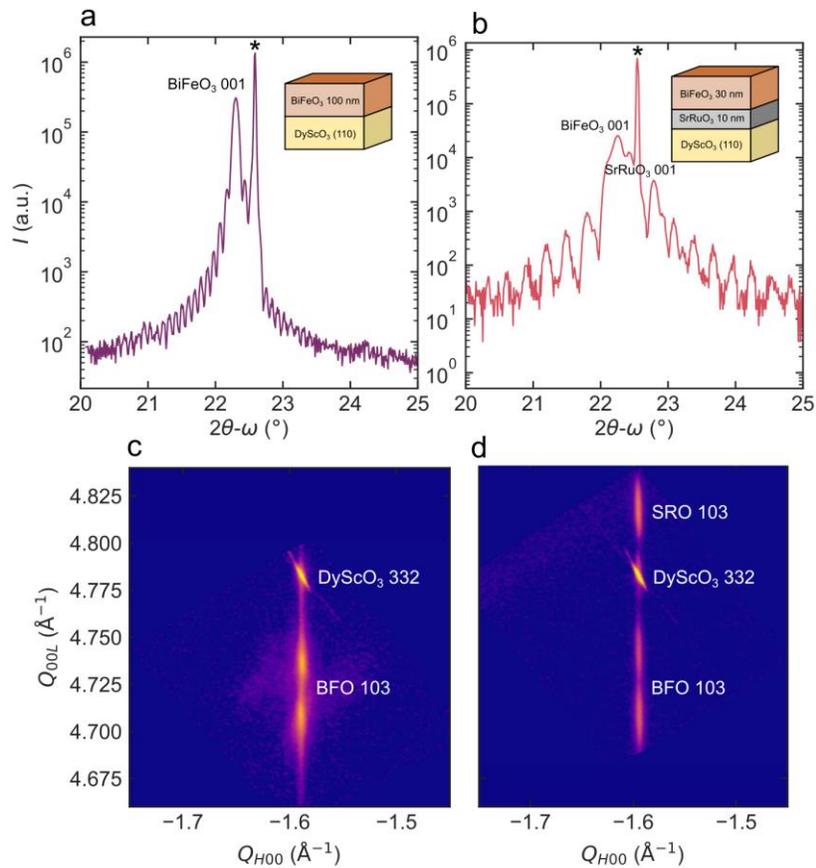

**Supp. Figure S2 | XRD of BiFeO₃ films.** Xray diffraction line scans of BFO thin films deposited directly on the substrate, **a**, and on a conducting SrRuO₃ back electrode, **b**, used in the experiments here. Pendellosung fringes confirm the quality and expected thickness of films. **c,d** Reciprocal space maps of the same films as a,b about the $103_{PC}/332_O$ diffraction peaks. $Q_{hkl}$ is labeled with respect to pseudocubic symmetry. In both cases, the films are epitaxially strained to the in-plane pseudocubic lattice constants of the DyScO₃ substrate, $a \cong b = 3.952$ Å. The splitting of the BFO 103 peaks in $Q_{00L}$ is due to the two rhombohedral variants from the two different ferroelectric domains[46]. Ferroelectric and magnetic domain structures in both samples, 100 nm and 30 nm, are approximately the same in the as-grown state, confirming their comparability.

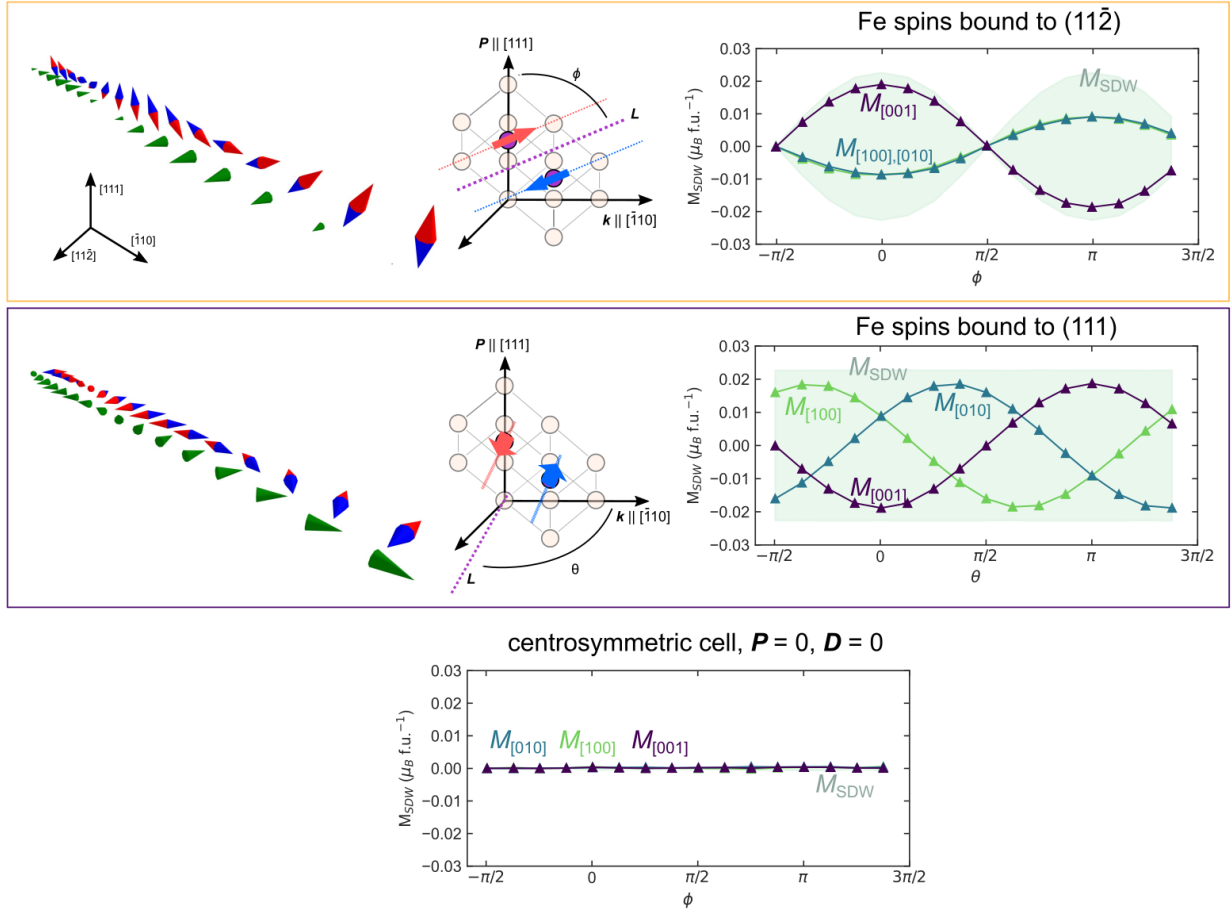

**Supp. Figure S3 | Modulation of $M_{SDW}$ due to the cycloid. a** In the discretized DFT calculations, when the Fe spins are rotated in the $(11\bar{2})$, we observe the emergence of $M_{SDW}$ in the $[11\bar{2}]$ direction due to the interaction with the DMI caused by the antiferrodistortive rotations of the $FeO_6$ octahedra, pointing along the direction of $P$. From symmetry, $H_{ij}^{DM} = D_{ij} \cdot (S_i \times S_j)$ which, because $S_i$ and $S_j$ are bound to the $(11\bar{2})$ plane and $D_{ij} \parallel P$, the energy is minimized when $S_{i,j} \parallel [\bar{1}10] \perp P$, thus $S_{i,j}$ cants in the $[11\bar{2}]$ direction. **b** We observe that when we bind the spins to other crystallographic planes, here (111), this interaction does not form a spin density wave and instead the resultant $M$ forms a cycloid with the same periodicity as the Fe spins. This unmodulated moment is thus compensated and would not be seen with NV microscopy[16], implying that the case in a must be approximately true in our samples. **c** As a further control, the same calculation is performed with an imposed centrosymmetry on the BFO unit cell. In this case, there is no canted moment from the simulations, confirming the mechanism of formation from $D_{ij}$ arising due to the ferroelectric polarization.

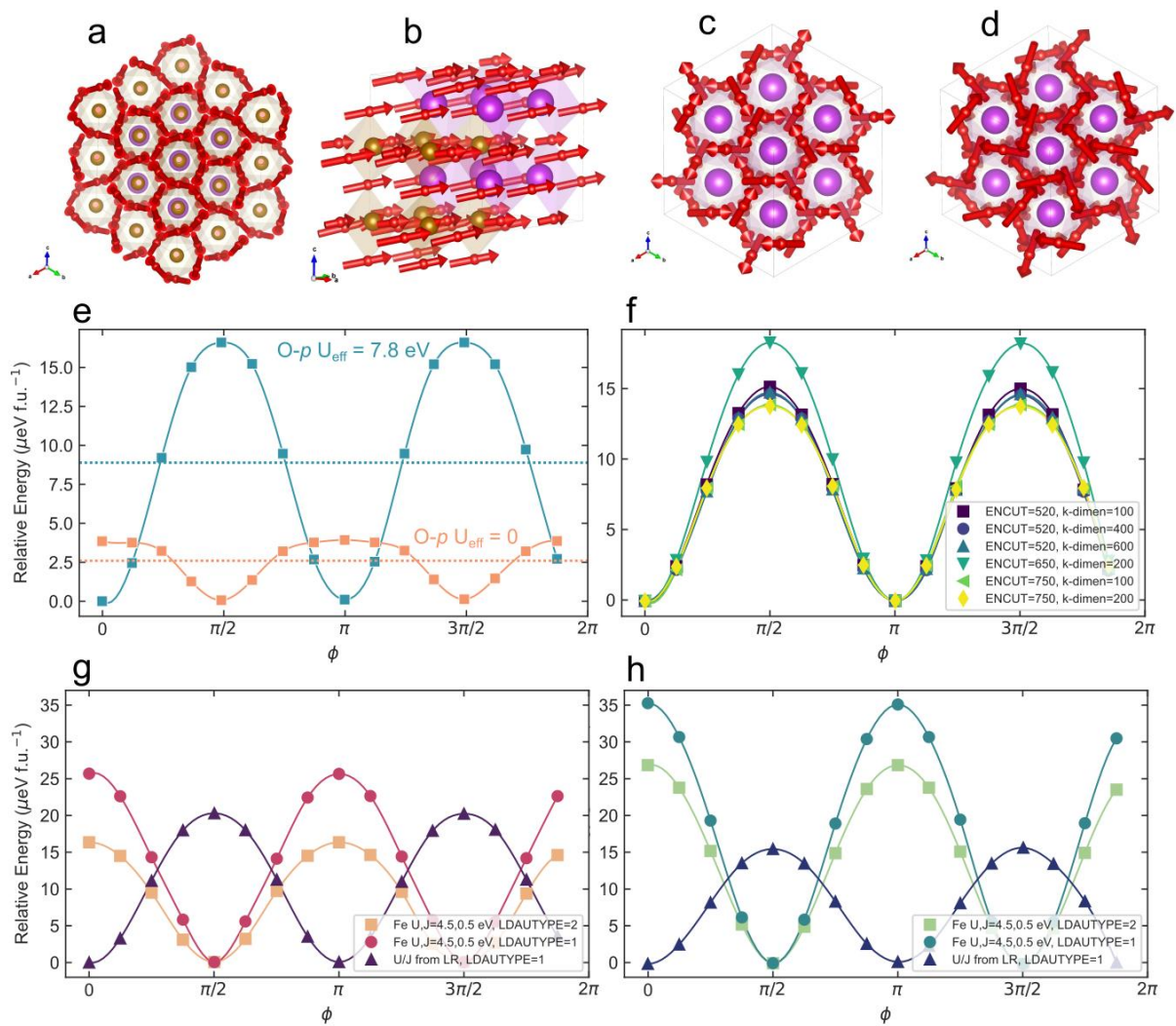

**Supp. Figure S4 | DFT methodology. a-d** Visualized distortions induced from various origins, as well as energy path comparisons over various DFT parameters. To provide a reference chirality of parent AFD distortions, **a,** arrows indicating the displacements of the cubic structure relative to the distorted structure, as reported in Ref. [18]. The relaxed relative distortions under strain is visualized in **b**. Additionally, **c** shows the reduction in antiferrodistortive (AFD) rotations due to the addition of Hund *J* applied to Fe-*d*, and **d**, the further reduction in AFD magnitude due to Hubbard corrections to the Bi-*p* and O-*p* manifolds. **e** Energies versus rotation angle for $\mathbf{k} \parallel [\bar{1}10]$, with and without $U_{\text{eff}}$ applied to the oxygen 2*p* states. The input structure itself is relaxed with Fe-*d*: *U,J* = 5.2, 0.4 eV and O-*p*: *U,J* = 9.7, 1.9 eV Hubbard corrections. **f** Anisotropy energy for different energy cutoffs (ENCUT) and *k*-point densities demonstrating that we have performed a proper convergence test versus the resolution of DFT calculations. **g,h** Energy versus rotation angle for

different on-site corrections, compared for two different relaxed structures. The linear response Hubbard corrections correspond to Fe-*d*: $U,J$ = 5.2,0.4 eV, Bi-s: $U,J$ = 0.8,0.8 eV, and O-*p*: $U,J$ = 9.7,1.9 eV. The structure in **g** is relaxed with these Hubbard parameters, whereas **h** is relaxed with $U_{eff}$ = 4.0 eV applied to Fe-*d* sites.

In the unstrained calculation, the on-site Hubbard *U* and Hund *J* parameters on O-*p*, in addition to Fe-*d*, are primarily responsible for the minimum in the anisotropy energy along [111], instead of the intersections with the (111) as previously reported[34]. For the *U*/*J* parameters listed above, we observe that using *U*/*J* applied to Fe-*d* states alone, the [111] corresponds to a local maximum in energy. Conversely, [111] is a local minimum if we apply on-site corrections to the O-*p* manifold. This result appears to be agnostic to the flavor of on-site correction used, i.e. DFT+*U*+*J*[47] versus DFT+$U_{eff}$[48] with $U_{eff}$ = *U*−*J*. This use of O-*p* Hubbard corrections could help to explain the differences between local minima on the MCAE energy landscape compared to previous DFT studies.

Additionally, when comparing DFT+*U*+*J* to DFT+$U_{eff}$, both an effective reduction of antiferrodistortive rotations of oxygen polyhedra, as well as a reduction of ferroelectric distortions of Fe atoms are observed. Conventionally, the inclusion of the Liechtenstein DFT+*U*+*J* formalism is known to amplify oxygen polyhedral distortions[49,50] compared to the Dudarev DFT+$U_{eff}$ counterpart. This can be explained on the basis that, in order to motivate $U_{eff}$, a spherical symmetry of the Coulomb exchange integrals is assumed[51] which is rarely the most accurate assumption, especially for transition metal oxide systems with symmetry broken by crystal field splitting[49,50]. In this case, however, the possible conflict between SOC and the Hund *J* antisymmetric intra-orbital exchange is also considered, as explored in reference [52] and [53].

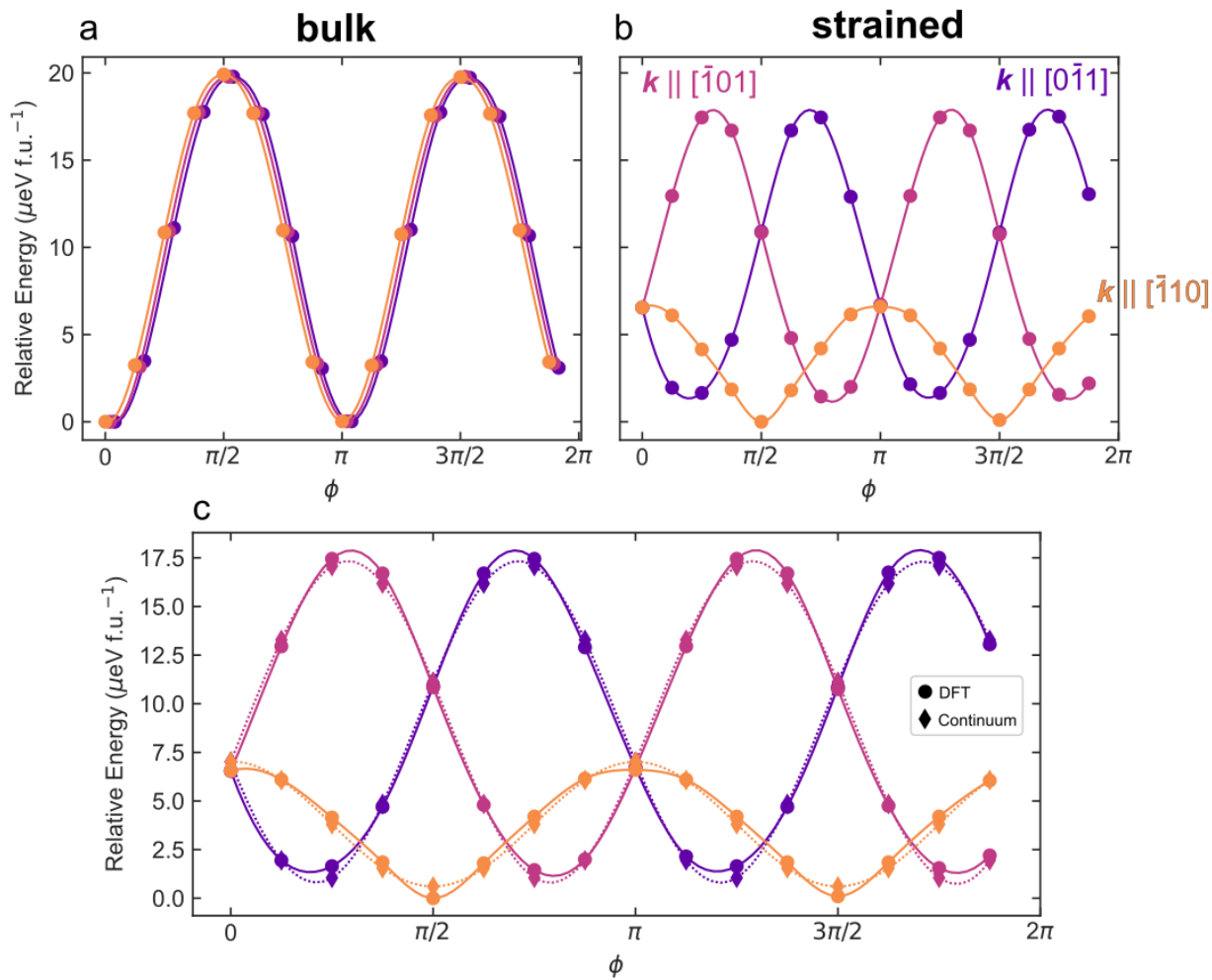

**Supp. Figure S5 | Strain dependence of the cycloid propagation.** Piecewise DFT calculations for the cycloid in both (**a**) epitaxially strained and (**b**) relaxed supercells, showing the dependence of $k$. The data in a is slightly offset in $\phi$ for visibility. In the bulk crystal, the three possible cycloid directions are energetically equivalent, which makes sense given the 3-fold symmetry of the rhombohedral unit cell. In the epitaxially strained case, this 3-fold symmetry is disrupted by the biaxial strain in the (001) due to epitaxy, inducing a preferred direction within the (001). **c** Data in b overlayed with fits to **Eq. S1** shown as dashed lines.

**Supp. Note S1 | Discussion of the magnetoelastic anisotropy.**

From the calculations presented here, the effective anisotropy is greatly reduced for the strained unit cell. This can be reconciled from the insight of previous DFT studies, which computationally confirmed that the single ion anisotropy is most strongly affected by ferroelectric distortions [X], while the net DMI vector is most sensitive to the magnitude of antiferrodistortive rotations of oxygen octahedra [X]. From **Supp. Figure S5**, strain induces a dominant cooperative distortion of oxygen, relative to iron and bismuth ions, which reduces the magnitude of the effective anisotropy. We observe the quantitative outcome of these distortions in the reduction of $|K_{eff}|$ in **Supp. Table S1**. This helps to rationalize, if the strained cell is not allowed to relax individual atomic positions, the experimentally observed behavior for the cycloid $k$ vector to orient along $[\bar{1}10]$, compared to $[\bar{1}01]$ and $[0\bar{1}1]$, is not reproduced.

In order to provide further evidence for the magnetoelastic anisotropy induced by epitaxial strain, the data here are also fit to the analytical expression for the energy versus rotation angle for BFO, derived in the SI of Ref. [54]. These results in **Supp. Figure S5** showcase an almost perfect agreement between the analytical expression and the data.

The anisotropy energy is defined in terms of the antiferromagnetic order parameter, oriented along the *z*-direction, which is parallel to [111]. $K_{eff}$ includes both uniaxial magnetocrystalline anisotropy (MAE), as well as DMI energy contributions[54]. The magnetostrictive term accounts for plane strain in (001), where the unit vector defines the direction normal to the film plane with **P** ∥ [111] under no applied stress. This energy can be expressed as:

$$F' = F_{anis} + F_{MS} = -K_{eff} L_z^2 - U(\boldsymbol{L} \cdot \hat{n}) \qquad (S1)$$

where $F'$ contains the effective anisotropy ($F_{anis}$) and plane-strain magnetostrictive ($F_{MS}$) contributions to the free energy and $\hat{n}$ is the unit vector that is normal to the thin film plane. Employing spherical coordinates, with the polar axis oriented along **P** ∥ [111] and the azimuthal axis along [11$\bar{2}$], we can express the AFM order parameter as $\boldsymbol{L} = |\boldsymbol{L}|[\sin\theta\cos\phi, \sin\theta\sin\phi, \cos\theta]$. Both energy contributions can then be expressed in terms of the polar and azimuthal angles $\theta$ and $\phi$:

$$F_{anis} = -K_{eff} \cos^2\theta \qquad (S2)$$
$$F_{MS} = -U(\sin\theta_n \sin\theta \cos\phi + \cos\theta_n \cos\theta)^2 \qquad (S3)$$

where $\theta_n$ defines the orientation of $\hat{n}$ within the polar reference frame. In the case of $\phi = \pi/2$, which corresponds to a $(11\bar{2})$ rotation plane,

$$F'_{11\bar{2}} = -(K_{eff} + U_{MS} \cos^2 \theta_n) \cos^2 \theta.$$

If $K_{eff} + U_{MS} \cos^2 \theta_n < 0$, the [111] axis is favored. If $K_{eff} + U_{MS} \cos^2 \theta_n > 0$, however, intersections between the $(11\bar{2})$ and $(111)$ planes (i.e. $[\bar{1}10]$ and $[1\bar{1}0]$) are preferred over the [111] axis. This agrees with the energy versus rotation angle provided in **Figure 2**.

**Supp. Table S1 | Fitted coefficients of Equation S1**

|  | $K_{eff}$ (μeV f.u.$^{-1}$) | $U_{MS}$ (μeV f.u.$^{-1}$) | $\theta_n$ |
|---|---|---|---|
| Unstrained | 19.655 | - | - |
| Strained | -3.011×10$^{-1}$ | -20.125 | 56.57° |

In **Supp. Table S1**, we report the fitted coefficients of the energy terms in **Eq. S1** for both the unstrained, relaxed structure, as well as the unit cell constrained to the DyScO$_3$ lattice parameters, $a \cong b \cong 0.394$ nm. The uncertainties of these obtained values from scipy.optimize.curve_fit, are less than 10$^{-14}$ eV for all $K_{eff}$ and $U_{MS}$ values, and 0.003° for $\theta_n$. By comparison, for the unstrained structure, $\theta_n = 54.7°$ [34]. The coefficient of the magnetostrictive term is negative, $U_{MS} < 0$, which is expected under compressive epitaxial strain, based on the intuition provided by Ref. [54]. According to the calculation presented here, $|U_{MS}| > |K_{eff}|$, which further supports the claim that in BiFeO$_3$, strain strongly affects the easy axis/plane preference and orientation.

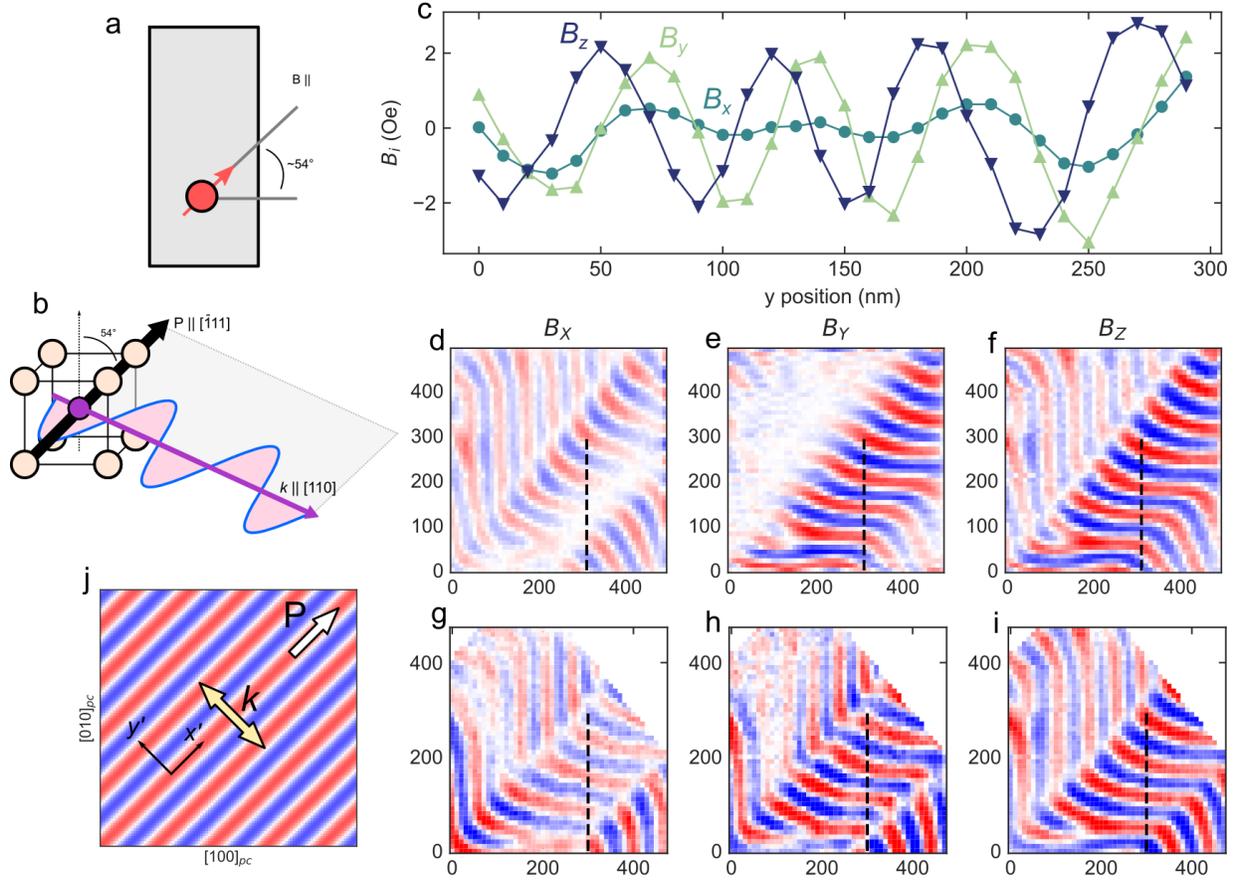

**Supp. Figure S6 | Components of B reconstructed from NV data. a** Schematic of the diamond tip used in NV microscopy, where the axis of the NV center points along the [111] direction of the diamond, approximately 54° from the surface. Using the analysis outlined in [43], the measured NV data can be used to reconstruct the *x*, *y*, and *z* components of *B* above the sample surface (**d,e,f**). Because the components of *B* are linearly dependent in Fourier space, they can be reconstructed from a single measurement of *B*∥ to the axis of the NV sensor. Axes are in nm. While this can be compared to a vector map of *B* reconstructed from three measurements at the same location of the sample (**g,h,i**), significant sources of error in the vector reconstruction such as uncertainties in the angle and location, as well as the fact that vector reconstruction takes three scans, make the analytical reconstruction preferable. From [16], the stray field from the spin cycloid in BFO on DSO can be expressed as:

$$B_x = A \sin(k(x-y))$$
$$B_y = -A \sin(k(x-y))$$
$$B_z = \sqrt{2} A \cos(k(x-y))$$

With the cycloid propagation vector $\vec{k}$ along [-110] in the lab frame when oriented with $x$ and $y$ parallel to the [100] and [010] pseudocubic crystallographic axes. If the reference frame is rotated 45°, such that $\vec{k} \parallel y'$ and $\vec{k} \perp x'$, where $y'$ and $x'$ are along [-110] and [110] respectively (illustrated in **j**), the expression for $B_i$ rotates such that:

$$B_{x'} = \frac{1}{\sqrt{2}}(B_x + B_y) = 0$$

$$B_{y'} = \frac{1}{\sqrt{2}}(B_x - B_y) = \sqrt{2}A\sin(k(x-y)) = \sqrt{2}A\sin(ky')$$

Here, we see that only $B_{y'}$, the field along the propagation vector $\vec{k}$, should be measurable, which is reflected in this reconstruction.

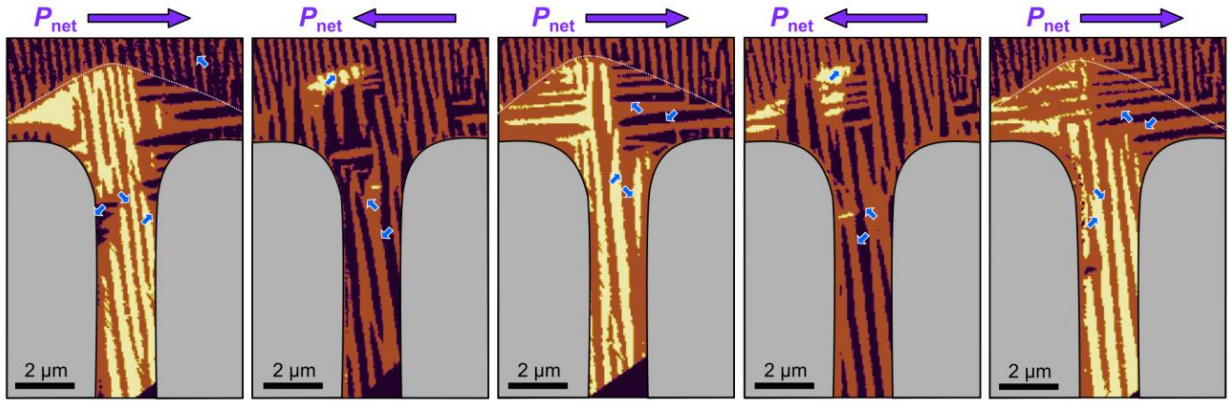

**Supp. Figure S7 | Ferroelectric domain switching with applied electric field.** Normalized PFM phase maps showing the reorientation of ferroelectric domains under successive electric fields. While ferroelectric domain walls remain generally stationary, with individual domains undergoing 71° in-plane switching, they are not completely so and may move with larger electric fields and longer pulse times.

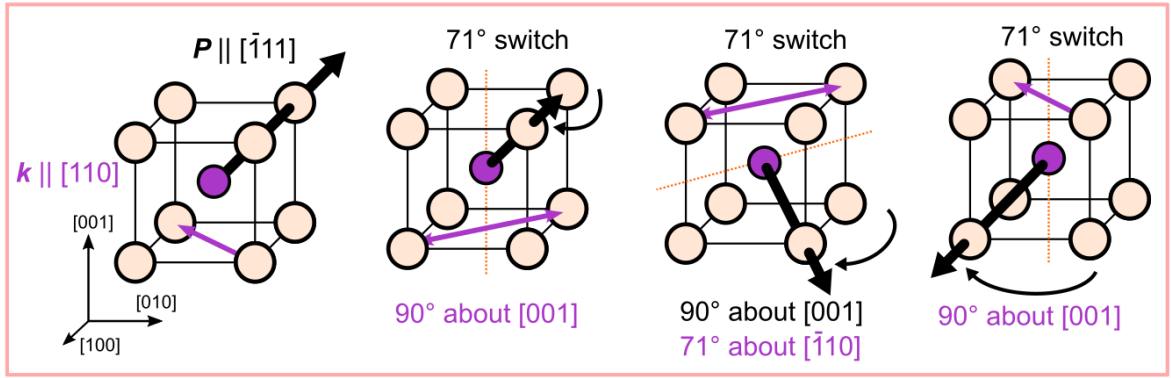
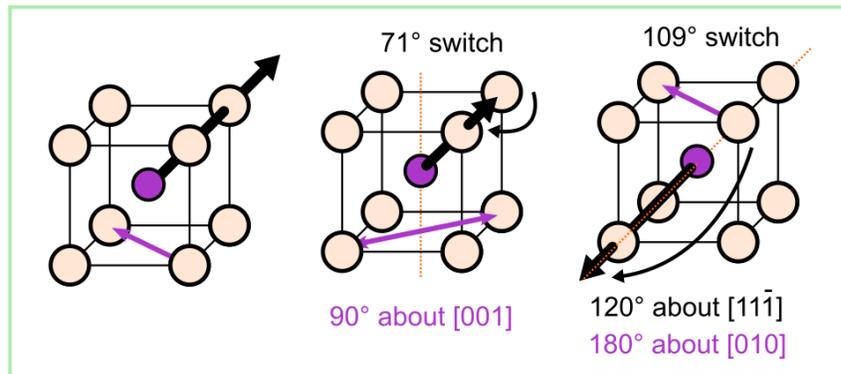
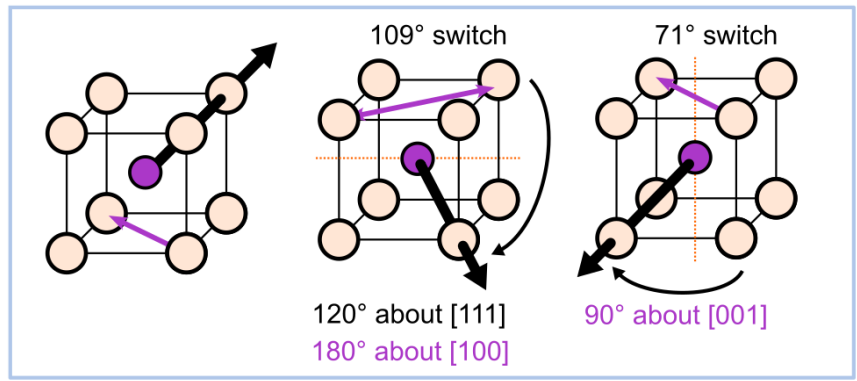

**Supp. Figure S8 | Switching pathways of $P$ and $k$.** Shown previously in BFO, in 180° ferroelectric switching events $P$ will rotate continuously through a combination of in-plane-71° and 109° switches. These three possible pathways with $P$ starting along $[\bar{1}11]$ are shown here. Operations shown in black are the higher symmetry operations that rotate $P$, but purple operations must be taken to preserve the experimentally observed sense of $k$. Because of the anisotropy in $k$, not all switching events can occur (in pseudocubic notation) through the highest symmetry operations, for example $C_3$ rotations through the high symmetry ⟨111⟩ axes, they must proceed through lower symmetry operations to preserve the in-plane nature of $k$.